\documentclass[aps, prd,preprint,showpacs,superscriptaddress,groupedaddress]{revtex4-1}
\usepackage{amsmath}
\usepackage[dvips]{graphicx}
\usepackage{morefloats}
\graphicspath{{Fig/}}
\begin{document}

\title{The late-time behaviour of tilted Bianchi type VIII universes in presence of diffusion}
\author{Dmitry Shogin}
\email{dmitry.shogin@uis.no}
\affiliation{Faculty of Science and Technology, University of Stavanger, N-4036 Stavanger, Norway}
\author{Sigbj\o rn Hervik}
\email{sigbjorn.hervik@uis.no}
\affiliation{Faculty of Science and Technology, University of Stavanger, N-4036 Stavanger, Norway}

\begin{abstract}
We apply the dynamical systems approach to ever-expanding Bianchi type~VIII cosmologies filled with a tilted~$\gamma$-fluid undergoing velocity diffusion on a scalar field. We determine the future attractors and investigate the late-time behaviour of the models. We find that at late times the normalized energy density~$\Omega$ tends to zero, while the scalar potential~$\Phi$ approaches~1 and dominates the evolution. Moreover, we demonstrate that in presence of diffusion fluids with~$\gamma<3/2$, which includes physically important cases of dust~${(\gamma=1)}$ and radiation~${(\gamma=4/3)}$, are asymptotically non-tilted; the velocity of the fluid with~$\gamma=3/2$ tends to a constant value~$0<\bar{V}<1$; and stiffer fluids evolve towards a state of extreme tilt. Finally, we show that diffusion significantly reduces the decay rates of energy density for dust and fluids stiffer than dust~$(\gamma \geq 1)$; for example, at~$\gamma=4/3$~(radiation) we obtain~${\rho/H^2 \propto e^{-3H_0 t}}$ at late times, while~$\rho/H^2 \propto e^{-4H_0 t}$ when diffusion is absent.
\end{abstract}


\maketitle

\section{Introduction}
The aim of theoretical cosmology is to investigate possible physical and mathematical properties of the universe. Anisotropic cosmological models, including the so-called Bianchi cosmologies \cite{Ellis2012,Groen2007,Christiansen2008} are of great theoretical importance and have been an object of close studies; see e.\,g.~\cite{Hewitt2001,Coley2008,Coley2005,Hervik2005,Collins1979,Hewitt1992, Harnett1996,Hervik2004,Coley2004,Hervik2006a}, where these models have been considered using a dynamical systems approach~\cite{Wainwright1997}. Special attention has been paid to determining the future asymptotic states of Bianchi cosmologies and describing the late-time dynamics of these universes. 
\par   
Type VIII universes are the most general ever-expanding Bianchi cosmologies and are therefore of special interest. These models have been studied before~\cite{Barrow2001,Ringstrom2000,Ringstrom2001,Ringstrom2003,Horwood2003}; in particular, the late-time dynamics of the model with a tilted $\gamma$-fluid~\cite{King1973} has been investigated in~\cite{Hervik2006}.
\par 
Models where a cosmological fluid is involved into diffusional interactions with a scalar field background can be considered as a generalization of models with a cosmological constant~\cite{Calogero2011,Calogero2012}. It has been shown, however, that models with diffusion can possess essentially different late-time dynamical properties~\cite{Shogin2013,Shogin2014}. Also, diffusion causes some interesting effects, which have been pointed out and studied numerically in FRW~\cite{Calogero2012,Calogero2013}, plane symmetric~$G_2$~\cite{Shogin2013} and Bianchi class~B~\cite{Shogin2014} universes.
\par 
In the current work we make a combined analytical and numerical analysis with a clear focus on the former. The aim of the paper is to confirm and generalize the results for Bianchi cosmologies obtained in~\cite{Shogin2014} and derive the expressions describing the asymptotic dynamical features of the universe in the general case of an arbitrary fully tilted~$\gamma$-fluid.
\par 
The paper is organized as follows. The evolution equations and constraints governing the dynamics of the model are given in section~\ref{Sec:Equations}. In section~\ref{Sec:Pre} we briefly describe the future asymptotic states of the model and also introduce some auxiliary variables. The obtained results are presented in sections~\ref{Sec:ResTilted} and~\ref{Sec:ResNonTilted} (the tilted and non-tilted case, respectively). Asymptotic dynamics of the geometric variables is discussed in section~\ref{Sec:ResGeom}. Summary is presented in section~\ref{Sec:Summary}.

\section{Equations of motion}
\label{Sec:Equations}
The Einstein field equations modified to take diffusion into account can be written as
\begin{equation}
R_{\alpha \beta}-\frac{1}{2}g_{\alpha \beta}+\phi g_{\alpha \beta} = T_{\alpha \beta}. \label{Eq:EqsCons:EFE}
\end{equation}
The diffusion equations are
\begin{align}
\nabla_\alpha T^{\alpha \beta} & =\nabla_\alpha (\phi g^{\alpha \beta})=DJ^\beta, \\
J^\alpha &= k\hat{u}^\alpha,\\
\nabla_\alpha J^\alpha &= 0, \label{Eq:EqsCons:Dif3}
\end{align}
in terms of notations introduced in~\cite{Shogin2014}~(see also~\cite{Calogero2011} for the background behind the diffusion model). We are using the orthonormal frame formalism~\cite{Elst1997} and write the equations (\ref{Eq:EqsCons:EFE})-(\ref{Eq:EqsCons:Dif3}) as a system of differential equations with algebraic constraints. The dimensionless time is introduced by
\begin{equation}
\frac{dt}{d\tau}=\frac{1}{H},
\end{equation}
where~$H$ is the Hubble scalar. The Hubble-normalized variables are then defined by
\begin{align}
(\Sigma_{ab}, N_{ab}, R_a) &= (\sigma_{ab}, n_{ab}, \Omega_a)/H, \\
(\Omega,\Phi) &= (\rho,\phi)/3H^2, \\
 K &= Dk/3H^3.
\end{align}
The normalized curvature~$N_{ab}$ and shear~$\Sigma_{ab}$ are parametrized by
\begin{align}
N_{ab} &=\sqrt{3}\left[ 
\begin{array}{ccc}
N_1 & 0 & 0 \\
0 & \bar{N}+N_- & N_{23} \\
0 & N_{23} & \bar{N}-N_-
\end{array} \right],\\
\Sigma_{ab} &=\left[ 
\begin{array}{ccc}
-2\Sigma_+ & \sqrt{3}\Sigma_{12} & \sqrt{3}\Sigma_{13} \\
\sqrt{3}\Sigma_{12} & \Sigma_+ +\sqrt{3}\Sigma_- & \sqrt{3}\Sigma_{23} \\
\sqrt{3}\Sigma_{13} & \sqrt{3}\Sigma_{23} & \Sigma_+-\sqrt{3}\Sigma_-
\end{array} \right].
\end{align} 
Using the approach proposed in~\cite{Coley2005,Hervik2006}, we introduce the complex variables
\begin{equation}
\begin{array}{ll}
\mathbf{N}_\times = N_-+iN_{23}, & \boldsymbol{\Sigma}_\times=\Sigma_-+i\Sigma_{23}, \\
\boldsymbol{\Sigma}_1=\Sigma_{12}+i\Sigma_{13}, & \mathbf{v}=v_2+iv_3.
\end{array}
\end{equation}
The remaining gauge freedom represents a complex rotation:
\begin{equation}
\phi: \quad [\mathbf{N}_\times,\boldsymbol{\Sigma}_\times,\boldsymbol{\Sigma_1},\mathbf{v}]\mapsto [e^{2i\phi}\mathbf{N}_\times,e^{2i\phi}\boldsymbol{\Sigma}_\times,e^{i\phi}\boldsymbol{\Sigma_1},e^{i\phi}\mathbf{v}].
\end{equation}
Following~\cite{Hervik2006}, we use the gauge function to replace~$R_1$ with~$\phi^\prime$ and then introduce ${R_3-iR_2=\sqrt{3}\mathbf{R}}$, where
\begin{equation}
\mathbf{R}=a\boldsymbol{\Sigma}_1+\mathbf{b}\boldsymbol{\Sigma}_1^*, \quad 
a=\frac{\bar{N}^2-N_1^2-\vert \mathbf{N}_\times \vert^2}{(\bar{N}-N_1)^2-\vert \mathbf{N}_\times \vert^2}, \quad \mathbf{b}=\frac{-2N_1\mathbf{N}_\times}{(\bar{N}-N_1)^2-\vert \mathbf{N}_\times \vert^2}.
\end{equation}
As mentioned in~\cite{Hervik2006}, both of these functions are bounded and therefore well-defined for tilted type~VIII models; namely,~$\vert a \vert \leq 1$ and~$\vert \mathbf{b} \vert \leq 1.$

The equations of motion are:
\begin{align}
\Sigma_+^\prime &=(q-2)\Sigma_++3\text{Re}(\mathbf{R}^*\boldsymbol{\Sigma}_1)-N_1\bar{N}-2\vert \mathbf{N}_\times \vert^2+\frac{\gamma \Omega}{2G_+}\left( -2v_1^2+\vert \mathbf{v} \vert^2 \right), \label{Eq:EqsCons:SigmaPlusEvol}\\
\boldsymbol{\Sigma}_\times^\prime &= (q-2+2i\phi^\prime)\boldsymbol{\Sigma}_\times+\sqrt{3}\boldsymbol{\Sigma}_1\mathbf{R}-\sqrt{3}\mathbf{N}_\times(2\bar{N}-N_1)+\frac{\sqrt{3}\gamma \Omega}{2G_+}\mathbf{v}^2,\\
\boldsymbol{\Sigma}_1^\prime &= 
(q-2+i\phi^\prime)\boldsymbol{\Sigma}_1-3\Sigma_+\mathbf{R}-\sqrt{3}\boldsymbol{\Sigma}_\times\mathbf{R}^*+\frac{\sqrt{3}\gamma\Omega v_1}{G_+}\mathbf{v},\\
\mathbf{N}_\times^\prime &= 
(q+2\Sigma_++2i\phi^\prime)\mathbf{N}_\times+2\sqrt{3}\boldsymbol{\Sigma}_\times\bar{N},\\
\bar{N}^\prime &=
(q+2\Sigma_+)\bar{N}+2\sqrt{3}\text{Re}(\boldsymbol{\Sigma}_\times^*\mathbf{N}_\times),\\
N_1^\prime &= (q-4\Sigma_+)N_1.
\end{align}
The equations for the fluid are:
\begin{align}
\Omega^\prime &= \frac{\Omega}{G_+} \left \{ 2q-(3\gamma-2)+ \left[ 2q(\gamma-1)-(2-\gamma) - \gamma \mathcal{S} \right] V^2 \right \} +\frac{K}{\sqrt{1-V^2}}, \label{Eq:EqsCons:OmegaEvol}\\
\Phi^\prime &= 2(q+1)\Phi-\frac{K}{\sqrt{1-V^2}},\\
v_1^\prime &= \left[T+2\Sigma_+-\frac{(\gamma-1)\sqrt{1-V^2}G_+}{\gamma G_-} \cdot \frac{K}{\Omega} \right]v_1-\sqrt{3}\text{Re}[(\mathbf{R}+\boldsymbol{\Sigma}_1)\mathbf{v}^*]+\sqrt{3}\text{Im}(\mathbf{N}_\times^*\mathbf{v}^2),\\
\mathbf{v}^\prime &= \left[T-\Sigma_+-i\sqrt{3}(\bar{N}-N_1)v_1+i\phi^\prime -\frac{(\gamma-1)\sqrt{1-V^2}G_+}{\gamma G_-} \cdot \frac{K}{\Omega}  \right]\mathbf{v} \nonumber \\
& \phantom{=}~~-\sqrt{3}(\boldsymbol{\Sigma}_\times+i\mathbf{N}_\times v_1)\mathbf{v}^*-\sqrt{3}(\boldsymbol{\Sigma}_1-\mathbf{R})v_1,\\
V^\prime &= \frac{V(1-V^2)}{G_-} \left[  (3\gamma-4)-\mathcal{S}-\frac{(\gamma-1)G_+}{\gamma \sqrt{1-V^2}} \cdot  \frac{K}{\Omega} \right], \label{Eq:EqsCons:VEvol}\\
K^\prime &= \frac{(\gamma-1)G_+V^2}{\gamma \sqrt{1-V^2} G_-}\cdot \frac{K^2}{\Omega}+\left[ 3q-\frac{V^2}{G_-}(3\gamma-4-\mathcal{S}) \right] K, \label{Eq:EqsCons:KEvol}
\end{align}
where
\begin{align}
q &= 2\Sigma^2+\frac{(3\gamma-2)+(2-\gamma)V^2}{2G_+}\Omega-\Phi,\\
V^2 &= v_1^2+\vert \mathbf{v} \vert^2, \\
\Sigma^2 &= \Sigma_+^2 +\vert \mathbf{\Sigma}_\times \vert^2+\vert \mathbf{\Sigma}_1 \vert^2, \\
\mathcal{S} &= \frac{1}{V^2}\Sigma_{ab}v^av^b, \\
T &= G_-^{-1}\left[ (3\gamma-4)(1-V^2)+(2-\gamma)V^2 \mathcal{S} \right],\\
G_\pm &= 1 \pm (\gamma-1)V^2.
\end{align}
The variables are subject to the following constraints:
\begin{align}
1 &= \Sigma^2+\frac{1}{4}N_1^2-N_1\bar{N}+\vert \mathbf{N}_\times \vert^2+\Omega+\Phi, \label{Eq:EqsCons:HamCons} \\
0 &= 2\text{Im}(\boldsymbol{\Sigma}_\times^*\mathbf{N}_\times)+\frac{\gamma \Omega v_1}{G_+}, \\
0 &= i\boldsymbol{\Sigma_1}(\bar{N}-N_1)+i\boldsymbol{\Sigma}_1^*\mathbf{N}_\times+\frac{\gamma \Omega}{G_+}\mathbf{v}. \label{Eq:EqsCons:Cons4}
\end{align}
 The state vector is considered to be
\begin{equation}
\text{X}=[\Sigma_+,\boldsymbol{\Sigma}_\times,\boldsymbol{\Sigma}_1,\mathbf{N}_\times, \bar{N},N_1,\Omega,\Phi,v_1,\mathbf{v},K]
\end{equation}
modulo the constraint equations (\ref{Eq:EqsCons:HamCons})--(\ref{Eq:EqsCons:Cons4}). The dimension of the physical state space is therefore eleven.
\par 
In the current paper we make an emphasis on analytical investigations of the system. However, numerical calculations have been used to check and confirm the obtained results. More precisely, we have performed the complete numerical integration of the system of evolution equations~(\ref{Eq:EqsCons:SigmaPlusEvol})--(\ref{Eq:EqsCons:KEvol}) for different sets of initial conditions. The constraints were used to control the errors and see when the numerical calculations break down (for numerical runs we have chose to use the 'F-gauge' given by~$\phi^\prime=0$, see~\cite{Coley2005} for a detailed discussion on different gauge choices).

\section{The future asymptotic states}
\label{Sec:Pre}
It was shown earlier in~\cite{Calogero2012,Shogin2013,Shogin2014} that cosmologies with diffusion may recollapse when the diffusion term is sufficiently large. Otherwise the models evolve towards the de Sitter state of accelerated expansion. The ever-expanding models are of particular interest, and we restrict our consideration to cosmologies of this kind.
\par 
Similarly to models with a positive cosmological constant, the future attractor for shear and curvature is 
\begin{equation}
[\Sigma_+,\boldsymbol{\Sigma}_\times,\boldsymbol{\Sigma}_1,\mathbf{N}_\times, \bar{N},N_1]=[0,0,0,0,0,0],
\end{equation}
the characteristic time scales of evolution being relatively small. This feature allows us to use a reduced system when investigating the stability of equilibrium sets for the fluid variables.
\par 
In ever-expanding cosmological models with diffusion both the energy density~$\Omega$ of the fluid and the diffusion term~$K$ are asymptotically zero, while the scalar potential~$\Phi\to 1$ and dominates the evolution at late times. Regarding the tilt, different situations~$V\to 0,$ $V\to \bar{V},$ $V\to 1$ are possible. To avoid mathematical difficulties, we introduce the following auxiliary variables:
\begin{equation}
X=\frac{K}{\Omega}
\end{equation}
for the case when the the fluid is not asymptotically extremely tilted, and
\begin{equation}
Y=\frac{K}{\Omega \sqrt{1-V^2}}, \qquad U=1-V^2
\end{equation}
otherwise. The evolution equations for these new variables are obtained from
equations~(\ref{Eq:EqsCons:OmegaEvol}) and~(\ref{Eq:EqsCons:VEvol}--\ref{Eq:EqsCons:KEvol}).
Also, it is convenient to introduce 
\begin{equation}
\Phi=1-f, \qquad V=\bar{V}+u, \qquad X=\bar{X}+x, \qquad Y=\bar{Y}+y
\end{equation}
to investigate the stability of the trivial equilibrium.

\section{Results for a fully tilted fluid}
\label{Sec:ResTilted}
\subsection{Case~$0<\gamma \leq 1$}
\label{Sec:Res:nonStiff}
The late-time attractor is given by
\begin{equation}
[\Omega,\Phi,V,K,X] = [0,1,0,0,0].
\end{equation}
The eigenvalues of the linearized system
\begin{equation}
[\lambda_\Omega,\lambda_f,\lambda_V,\lambda_K,\lambda_X]=[-3\gamma,-2,3\gamma-4,-3,3\gamma-3]
\end{equation}
are all real and negative at~$0<\gamma<1$. At~$\gamma=1$ one obtains~$\lambda_X=0$; however, a simple analysis shows that the equilibrium is stable and the leading-order approximations are (hatted variables are certain constants):
\begin{align}
\Omega &\approx \left\{ \begin{array}{ll} 
\hat{\Omega}e^{-3\gamma \tau}, & 0<\gamma <1; \\
\hat{K}\tau e^{-3\tau}, &	\gamma=1; \end{array}\right. \\
\Phi &\approx \left\{ \begin{array}{ll} 
1-\hat{\Phi}e^{-3\gamma \tau}, & 0<\gamma<2/3; \\
1-\hat{\Phi}e^{-2\tau}, & 2/3\leq \gamma \leq 1;
\end{array}\right. \\
V &\approx \hat{V}e^{-(4-3\gamma)\tau};\\
K &\approx \hat{K}e^{-3\tau};\\
X &\approx \left\{ \begin{array}{ll} 
\hat{X}e^{-3(1-\gamma) \tau}, & 0<\gamma<1; \\
\displaystyle{\frac{\hat{X}}{\tau-\tau_0}}, &	\gamma=1. \end{array}\right.
\end{align}
In practice, however, it is necessary to take the higher-order terms into account to get an accurate approximation for the late-time expressions for~$\Omega$ and~$\Phi$. The role of higher-order terms is largest in the following special ranges: at~$2/3<\gamma \leq 1$ for the potential, and in the proximity of~$\gamma=1$ for the energy density. A closer investigation yields:
\begin{align}
\Omega &\approx \left\{ \begin{array}{ll} 
\displaystyle{\hat{\Omega}e^{-3\gamma \tau}-\frac{\hat{K}}{3(1-\gamma)}e^{-3\tau}}, & \gamma \text{~close to~}1;\\
\hat{K}\tau e^{-3\tau}+\hat{\Omega}e^{-3\tau}; & \gamma=1; \end{array} \right.\\
\Phi &\approx \left\{ \begin{array}{ll} 
\displaystyle{1-\hat{\Phi}e^{-2\tau}-\hat{\Omega}e^{-3\gamma \tau}+\frac{\hat{K}}{3(1-\gamma)}e^{-3\tau}}, & 2/3<\gamma<1; \\
1-\hat{\Phi}e^{-2\tau}-\hat{K}\tau e^{-3\tau}-\hat{\Omega}e^{-3\tau}, & \gamma=1. \end{array}\right.
\end{align}
\subsection{Case $1<\gamma<3/2$}
\label{Sec:Res:PreRad}
The attractor is
\begin{equation}
[\Omega,\Phi,V,K,X] = [0,1,0,0,3(\gamma-1)].
\end{equation}
All the eigenvalues of the linearized system are real and negative in the given range of~$\gamma$:
\begin{equation}
[\lambda_\Omega,\lambda_f,\lambda_V,\lambda_K,\lambda_x]=\left[-3,-2,2-\frac{3}{\gamma},-3,3-3\gamma\right].
\end{equation}
The corresponding leading-order approximation is:
\begin{align}
\Omega &\approx \hat{\Omega}e^{-3\tau}; \\
\Phi 	&\approx 1-\hat{\Phi}e^{-2\tau}; \\
V &\approx \hat{V}e^{-(\frac{3}{\gamma}-2)\tau};\\
K &\approx \hat{K}e^{-3\tau};\\
X &\approx 3(\gamma-1)+\hat{X}e^{-3(\gamma-1)\tau}.
\end{align}
Similarly to the case~$0<\gamma \leq 1,$ considering higher-order terms is necessary to obtain accurate approximations for~$\Omega$ and~$\Phi$. The more precise asymptotic expressions valid for the values of~$\gamma$ close to~$\gamma=1$ are:
\begin{align}
\Omega &\approx \frac{\hat{K}}{3(\gamma-1)}e^{-3\tau}+\hat{\Omega}e^{-3\gamma \tau};\\
\Phi &\approx  
1-\hat{\Phi}e^{-2\tau}-\frac{\hat{K}}{3(\gamma-1)}e^{-3\tau}-\hat{\Omega}e^{-3\gamma \tau}.
\end{align}

\subsection{Case $\gamma=3/2$}
\label{Sec:Res:Trans}
In this case the fluid is asymptotically tilted, and the attractor is
\begin{align}
[\Omega,\Phi,V,K,X] &= [0,1,\bar{V},0,\bar{X}],\\
\text{with}~~\bar{X} &= \frac{3\sqrt{1-{\bar{V}}^2}}{2+{\bar{V}}^2}.
\end{align}
The eigenvalues of the linearized system are
\begin{equation}
[\lambda_\Omega,\lambda_f,\lambda_v,\lambda_K,\lambda_x]=\left[-3,-2,0,-3,-\frac{3-2{\bar{V}}^2}{2-{\bar{V}}^2}\right].
\end{equation}
A more detailed analysis confirms the stability of the equilibrium set with the late-time dynamics described by
\begin{align}
\Omega &\approx \hat{\Omega}e^{-3\tau}; \\
\Phi		&\approx 1-\hat{\Phi}e^{-2\tau};\\
V			&\approx \bar{V}+\hat{V}e^{-\frac{3-2{\bar{V}}^2}{2-{\bar{V}}^2}\tau};\\
K			&\approx \hat{K}e^{-3\tau}; \\
X			&\approx \bar{X}+\hat{X}e^{-\frac{3-2{\bar{V}}^2}{2-{\bar{V}}^2}\tau};
\end{align}
In contrast to the previously considered cases, the higher-order terms do not play a substantial role.

\subsection{Case $3/2<\gamma<2$}
\label{Sec:Res:Stiff}
The attractor is
\begin{equation}
[\Omega,\Phi,U,K,Y]=[0,1,0,0,1],
\end{equation}
where~$U\to 0$ implies~$V\to 1:$ the tilt is asymptotically extreme. The corresponding eigenvalues of the linearized system are real and negative at~$3/2<\gamma<2$:
\begin{equation}
[\lambda_\Omega,\lambda_\Phi,\lambda_U,\lambda_K,\lambda_Y]=\left[-3,-2,-\frac{4\gamma-6}{2-\gamma},-\frac{3-\gamma}{2-\gamma},-1\right].
\end{equation}
The solution possesses a special feature in this range of~$\gamma$; namely, the fluid velocity is a slowly-varied function and cannot be treated as small with respect to constants. As a result, the asymptotic regime arrives at extremely late times; the expressions one obtains cannot serve as approximations at physical times. The dynamics of the variables can be then characterized as exponential decay with tilt-dependent non-constant exponents. An exception is the scalar potential, for which the expression
\begin{equation}
\Phi \approx 1-\hat{\Phi}e^{-2\tau}
\end{equation}
still gives an appropriate description of the late-time behaviour.

\section{Results for a non-tilted fluid}
\label{Sec:ResNonTilted}
The attractor is
\begin{equation}
[\Omega,\Phi,K]=(0,1,0).
\end{equation}
The fluid velocity does not affect the leading terms at~$0<\gamma \leq 3/2$, so the expressions for the tilted fluid obtained in sections~\ref{Sec:Res:nonStiff}--\ref{Sec:Res:Trans} are also valid in the non-tilted case. However,~$\gamma=3/2$ ceases to be a special point when the fluid velocity is identically zero. The working approximation for~$3/2\leq \gamma <2$ by this becomes
\begin{align}
\Omega &\approx \hat{\Omega}e^{-3\tau}; \\
\Phi &\approx 1-\hat{\Phi}e^{-2\tau}; \\
K &\approx \hat{K}e^{-3\tau}.
\end{align}

\section{On the late-time behaviour of the geometric variables}
\label{Sec:ResGeom}
For general ever-expanding type~VIII cosmologies with diffusion, the $\Omega$-containing terms do not affect the leading-order approximations at~$0<\gamma<3/2$. The future asymptotic behaviour of the geometric variables in this range is given by
\begin{align}
[\Sigma_+,\boldsymbol{\Sigma}_\times] &\approx [\hat{\Sigma}_+,\hat{\boldsymbol{\Sigma}}_\times]\cdot e^{-2\tau}; \label{Eq:Res:SigmaApprox}\\
\boldsymbol{\Sigma}_1 &\approx \hat{\boldsymbol{\Sigma}}_1 e^{-3\tau};\label{Eq:Res:Sigma123Approx}\\
[\mathbf{N}_\times, \bar{N},N_1] &\approx [\hat{\mathbf{N}}_\times, \hat{\bar{N}},\hat{N_1}]\cdot e^{-\tau}\label{Eq:Res:NApprox}.
\end{align}
However, at larger values of~$\gamma$ the~$\Omega$-term plays an important role in future asymptotics of~$\boldsymbol{\Sigma}_1$. An accurate study shows that at~$\gamma=3/2$ the relation~(\ref{Eq:Res:Sigma123Approx}) should be replaced by
\begin{equation}
\boldsymbol{\Sigma}_1 \approx \frac{3\sqrt{3}\hat{\Omega}\bar{v}_1\bar{\mathbf{v}}}{2+{\bar{V}}^2}\tau e^{-3\tau}+\hat{\boldsymbol{\Sigma}}_1e^{-3\tau},
\end{equation}
while~(\ref{Eq:Res:SigmaApprox}) and~(\ref{Eq:Res:NApprox}) still hold. At~$3/2<\gamma<2$ the situation becomes similar to that described in section~\ref{Sec:Res:Stiff}. The shear~$\boldsymbol{\Sigma}_1$ decays approximately exponentially, but due to the slowly-alterating tilt the exponent cannot be treated as a constant. The asymptotic regime does not arrive at physical times.
\par 
Note that said above is valid only for~{\it fully} tilted models. For one-tilted~($\mathbf{v}=0$) and non-tilted~($V=0$) cosmologies the~$\Omega$-term in evolution equation for~$\mathbf{\Sigma}_1$ is identically zero, and, therefore, expressions~(\ref{Eq:Res:SigmaApprox})--(\ref{Eq:Res:NApprox}) give a valid approximation in the whole range~$0<\gamma<2$.

\section{Summary}
\label{Sec:Summary}
The performed investigation shows that although cosmologies with diffusion are in some aspects similar to those with a positive cosmological constant, the late-time behaviour of these models can be substantially different. 
\par 
The key result is that in presence of diffusion all fluids with~$\gamma<3/2$ are asymptotically non-tilted. This range includes the cases of largest physical importance, namely dust~($\gamma=1$) and radiation~($\gamma=4/3$). On the contrary, in the absence of diffusion only fluids less stiff than radiation isotropize; radiation itself remains tilted in far future.
\par 
In cases when the fluid asymptotically becomes non-tilted the tilt decreases exponentially, both with and without diffusion. However, the decay rates are equal only for dust and fluids less stiff than dust~$(\gamma\leq 1).$ The velocity of fluids stiffer than dust decreases faster when diffusion is present; at late times
\begin{equation}
V \propto \left\{ \begin{array}{lll}
e^{-(4-3\gamma)\tau}, & 1<\gamma<4/3, & \text{without diffusion};\\
e^{-\left( \frac{3}{\gamma}-2 \right)\tau}, & 1<\gamma<3/2, & \text{with diffusion}.
\end{array} \right. 
\end{equation}
\par 
Both in diffusive and diffusionless case the universes evolve towards a vacuum state. However, in presence of diffusion the asymptotic decay rates of energy density are substantially reduced for fluids stiffer than dust:
\begin{equation}
\Omega \propto \left\{ \begin{array}{lll}
e^{-3\gamma \tau}, & 1\leq \gamma\leq 4/3, & \text{without diffusion};\\
e^{-3\tau},			& 1<\gamma \leq 3/2, & \text{with diffusion};
\end{array} \right.
\end{equation}
Dust becomes a special case when diffusion is present, the late-time dynamics of its energy density being approximated by
\begin{equation}
\Omega \approx \hat{K}\tau e^{-3\tau}+\hat{\Omega}e^{-3\tau}.
\end{equation}
\par 
Regarding the scalar potential, diffusion does not affect the leading-order terms in the asymptotic expression. However, the higher-order terms play an important role in constructing a valid approximation in the proximity of~$\gamma=1$; the impact of diffusion on~$\Phi$ is thus substantial in this range.
\par 
The diffusion term itself decays exponentially in future; namely, at later times~$K\propto e^{-3\tau}$ in the range~$0<\gamma \leq 3/2$.
\par 
Finally, we should make an important remark. The characteristic time scales of evolution are substantially extended for cosmologies passing close to the point of recollapsation. This happens when the diffusion term is initially large enough~\cite{Shogin2013,Shogin2014}. This means that the asymptotic regime does not arrive at physical times and the approximations obtained in this paper cannot in practice be applied to such universes. 

\bibliography{Bib/New}
\end{document}